\documentclass{article}
\usepackage{ijcai20}
\usepackage{color}
\usepackage{times}
\usepackage{soul}
\usepackage{url}
\usepackage[hidelinks]{hyperref}
\usepackage[utf8]{inputenc}
\usepackage[small]{caption}
\usepackage{graphicx}
\usepackage{amsmath}
\usepackage{amsthm}
\usepackage{booktabs,bigstrut}
\usepackage{algorithm}
\usepackage{algorithmic}
\title{DIDFuse: Deep Image Decomposition for Infrared and Visible Image Fusion}

\author{
Zixiang Zhao$^1$\and
Shuang Xu$^1$\footnote{Co-first Author}\and
Chunxia Zhang$^1$\and
Junmin Liu$^1$\and
Jiangshe Zhang$^1$\footnote{Corresponding Author}\And
Pengfei Li$^2$
\\
\affiliations
$^1$School of Mathematics and Statistics, Xi'an Jiaotong University, China\\
$^2$Hikvision, China\\
\emails
\{zixiangzhao, shuangxu\}@stu.xjtu.edu.cn,
\{cxzhang, junminliu, jszhang\}@mail.xjtu.edu.cn,
lipengfei27@hikvision.com
}
\graphicspath{{Figure/}}
\usepackage{multirow}
\usepackage{multicol}
\usepackage[american]{babel}
\usepackage{microtype}
\begin{document}

\maketitle

\begin{abstract}
Infrared and visible image fusion, a hot topic in the field of image processing, aims at obtaining fused images keeping the advantages of source images. This paper proposes a novel auto-encoder~(AE) based fusion network. The core idea is that the encoder decomposes an image into background and detail feature maps with low- and high-frequency information, respectively, and that the decoder recovers the original image. To this end, the loss function makes the background/detail feature maps of source images similar/dissimilar. In the test phase, background and detail feature maps are respectively merged via a fusion module, and the fused image is recovered by the decoder. Qualitative and quantitative results illustrate that our method can generate fusion images containing highlighted targets and abundant detail texture information with strong reproducibility and meanwhile surpass state-of-the-art (SOTA) approaches.
\end{abstract}

\section{Introduction}
Image fusion is an image processing technique for information enhancement. The principle is to preserve the complementary and redundant information from source images containing the same scene without artifacts \cite{meher2019a}.
In image fusion, the infrared and visible image fusion, a.k.a. IVIF, can be applied to many domains, such as surveillance  \cite{bhatnagar2015novel}, modern military and fire rescue tasks \cite{lahoud2018ar,hu2017adaptive}, face recognition \cite{ma2016infrared} and so on.

Proverbially, infrared images can avoid visually cognitive obstacles caused by illumination changes and artifacts, but they are with low spatial resolution and poor texture detail information. Conversely, visible images are with high spatial resolution and rich information of appearance and gradient, while they are easily affected by obstructions and light reflections. Therefore, making the fusion image retain both thermal radiation information of the infrared images and gradient information of the visible images will be conducive to target recognition and tracking.

In general, the IVIF algorithms can be divided into two groups: traditional methods and deep learning methods. Specifically, representative traditional methods include image multi-scale transformation \cite{li2011performance}, sparse representation \cite{zong2017medical}, subspace learning \cite{patil2011image} and
the saliency based method \cite{zhang2017infrared}.

Currently, deep learning (DL) has emerged as a prevalent tool in the field of IVIF. DL based methods can be categorized into three groups. The first group is based on Generative Adversarial Networks (GANs). In FusionGAN \cite{ma2019fusiongan}, a generator creates fused images with infrared thermal radiation and visible gradient information and a discriminator forces the fused images to have more details from the visible images. In the light of Conditional GANs \cite{mirza2014conditional}, detail preserving GAN \cite{ma2020infrared} changes the loss function of FusionGAN for improving the quality of detail information and sharpening the target boundary. The second group \cite{li2018infrared,lahoud2019fast} is an extension of image multi-scale transformation. Generally speaking, they transform images from the spatial domain to background and detail domains by means of filters or optimization based methods. Background images are simply averaged. Since there are high-frequency textures in detail images, they fuse feature maps of detail images extracted from a pre-trained network (for example, VGG \cite{simonyan2014very}). At last, a fusion image is recovered by merging the fused background and detail images. The third group consists of AE based methods \cite{li2018densefuse}. In the training phase, an AE network is trained. In the test phase, they fuse feature maps of source images, which then pass through the decoder to recover a fusion image. In summary, in DL based methods, deep neural networks (DNNs) are often employed to extract features of input images and then a certain fusion strategy is exploited to combine features to complete the image fusion task.

It is worth pointing out a shortcoming of the second group, i.e. DL is used only in the fusion stage and they employ filters or optimization based methods in the decomposition stage. To overcome this shortcoming, by combining principles of the second and third groups, we propose a novel IVIF network, called deep image decomposition based IVIF (DIDFuse). Our contributions are two-fold:

(1) To the best of our knowledge, this is the first deep image decomposition model for IVIF task, where both fusion and decomposition are accomplished via an AE network. The encoder and the decoder are responsible for image decomposition and reconstruction, respectively. In training phase, for decomposition stage, the loss function forces background and detail feature maps of two source images similar/dissimilar. Simultaneously, for reconstruction stage, the loss function maintains pixel intensities between source and reconstructed images, and gradient details of the visible image. In the test phase, background and detail feature maps of test pairs are separately fused according to a specific fusion strategy, and then the fused image can be acquired through the decoder.

(2) As far as we know, the performance of existing IVIF methods \cite{ma2016infrared,li2018densefuse,zhang2017infrared,li2018infrared} is only verified on a limited number of hand-picked examples in TNO dataset.  However, we test our model on three datasets, including TNO, FLIR and NIR. In total, there are 132 test images with indoor and outdoor scenes, and with daylight and nightlight illuminations. Compared with SOTA methods, our method can robustly create fusion images with brighter targets and richer details. It can be potentially utilized in target recognition and tracking.

The remaining article is arranged as follows. Related work is introduced in section \ref{sec:2}. The mechanism of the proposed network is described in section \ref{sec:3}. Then, experimental results are reported in section \ref{sec:4}. At last, some conclusions are drawn in section \ref{sec:5}.

\section{Related Work} \label{sec:2}
Since our network structure is closely related with U-Net, we introduce U-Net in section \ref{sec:2_1}. Then, traditional two-scale image decomposition methods are briefly reviewed in section \ref{ts_de}.
\subsection{U-Net and Skip Connection}\label{sec:2_1}
U-Net is applied to biomedical image segmentation\cite{ronneberger2015u}, similar to AE network, U-Net consists of a contracting path for feature extraction and an expanding path for precise localization. Compared with AE, there is a channel-wise concatenation of corresponding feature maps from contracting and expanding paths in U-Net. In this manner, it can extract “thicker” features that help preserve image texture details during downsampling. In literature \cite{mao2016image}, a U-Net-like symmetric network is used for image restoration. It employs skip connection technique, where feature maps of convolution layers are added to corresponding deconvolution layers to enhance the information extraction capability of the neural network and to accelerate convergence.
\subsection{Two-Scale Decomposition}\label{ts_de}
As a subset of multi-scale transformation, two-scale decomposition in IVIF decomposes an original image into background and detail images with background and target information, respectively.
In \cite{li2018densefuse}, given an image $I$, they obtained the background image $I^{b}$ by solving the following optimization problem,
$$
I^{ b} = \arg\min ||I-I^{ b}||_F^2+\lambda (||g_x*I^{ b}||_F^2+||g_y*I^{ b}||_F^2),$$
where $*$ denotes a convolution operator, and $g_x=[-1,1]$ and $g_y=[-1,1]^T$ are gradient kernels. Then, the detail image is acquired by $I^{d}=I-I^{b}$. Similarly, a box filter is used to get the background image in \cite{lahoud2019fast}, and the method of obtaining the detail image is the same as that of \cite{li2018densefuse}. After decomposition, background and detail images are separately fused with different criteria. At last, the fused image is reconstructed  by combining fused background and detail images.

\section{Method}\label{sec:3}
In this section, we will introduce our DIDFuse algorithm and the proposed network structure. In addition, details of training and testing phases are also illustrated.

\subsection{Motivation}
As described in section \ref{ts_de}, two-scale decomposition decomposes the input image into a background image containing low-frequency information with large-scale pixel intensity changes and a detail image embodying high-frequency information with small-scale pixel intensity changes.
Currently, most algorithms incorporate certain prior knowledge, and employ filters or optimization based methods to decompose images. Hence, they are manually designed decomposition algorithms. We highlight that image decomposition algorithms are intrinsically feature extractors. Formally, they transform source images from spatial domain into feature domain. It is well known that the DNN is a promising data-driven feature extractor and has great superiority over traditional manually-designed methods. Unfortunately, it lacks a DL based image decomposition algorithm for IVIF task.

Consequently, we present a novel deep image decomposition network in which an encoder is exploited to perform two-scale decomposition and extract different types of information, and a decoder is used to recover original images.

\begin{figure*}[t]
	\centering
	\includegraphics[width=\linewidth]{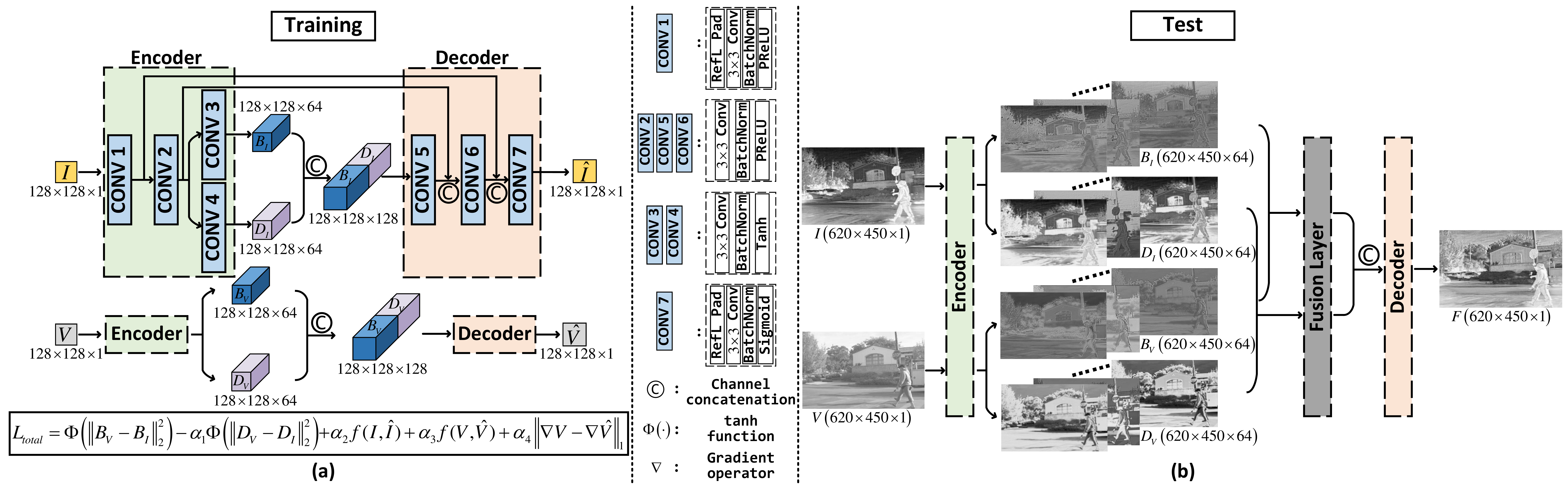}
	\caption{Neural network framework of DIDFuse.}
	\label{DIDF}
\end{figure*}

\subsection{Network Architecture}
Our neural network consists of an encoder and a decoder. As illustrated in Figure \ref{DIDF}, the encoder is fed with an infrared or a visible image and generates background and detail feature maps. Then, the network concatenates two kinds of feature maps along channels. At last, concatenated feature maps pass through decoder to recover the original image. To prevent the detail information of the feature maps from being lost after multiple convolutions and to speed up the convergence, we add the feature maps from the first and second convolutions to the inputs of the last and penultimate convolutions, and the adding strategy is concatenating the corresponding feature maps along channels. As a consequence, the pixel intensity and gradient information of the source images can be better retained in the reconstructed image.

\begin{table}[tbp]
	\centering
	\begin{tabular}{cccccc}
		\toprule
		Layers & Size & InC & OutC & Padding & Activation\\
        \midrule
		conv1 & 3     & 1     & 64    & Reflection & PReLU\\
		conv2 & 3     & 64    & 64    & Zero  & PReLU \\
		conv3 & 3     & 64    & 64    & Zero  & Tanh \\
		conv4 & 3     & 64    & 64    & Zero  & Tanh \\
		conv5 & 3     & 128   & 64    & Zero  & PReLU \\
		conv6 & 3     & 64    & 64    & Zero  & PReLU \\
		conv7 & 3     & 64    & 1     & Reflection & Sigmoid\\
		\bottomrule		
	\end{tabular}%
	\caption{Network configuration. Size denotes the size of convolutional kernel. InC and OutC are the numbers of input and output channels, respectively.}
	\label{tab:net_configuration}%
\end{table}%

Table \ref{tab:net_configuration} lists the network configuration. Encoder and decoder contain four and three convolutional layers, respectively. Each layer consists of a padding, a $3\times3$ convolution, a batch normalization and an activation function. The first and the last layers utilize reflection padding to prevent artifacts at the edges of the fused image. Activation functions of \texttt{conv3} and \texttt{conv4} are set to the hyperbolic tangent function (tanh) since they output background and detail feature maps. As for \texttt{conv7}, it is activated by sigmoid function since it reconstructs original images. Other layers are followed by parametric rectified linear units (PReLU).

\subsection{Loss Function}
In the training phase, we aim to obtain an encoder that performs two-scale decomposition on the source images, and at the same time, acquire a decoder that can fuse the images and preserve the information of source images well. The training process is shown in Figure \ref{DIDF}(a).

\paragraph{Image decomposition.}Background feature maps are used to extract the common features of source images, while detail feature maps are used to capture the distinct characteristics from infrared and visible images. Therefore, we should make the gap of background feature maps small. In contrast, the gap of detail feature maps should be great. To this end, the loss function of image decomposition is defined as follow,
\begin{equation}\label{L1}
{L_1} = \Phi \left( {{{\left\| {{B_V} - {B_I}} \right\|}_2^2}} \right) - {\alpha _1}\Phi \left( {{{\left\| {{D_V} - {D_I}} \right\|}_2^2}} \right),
\end{equation}
where $B_V$, $D_V$ are the background and detail feature maps of the visible image $V$, and $B_I$, $D_I$ are those of the infrared image $I$. $\Phi \left(  \cdot  \right)$ is tanh function that is used to bound gap into interval $(-1,1)$.

\subsubsection{Image Reconstruction}
As for image reconstruction, to successfully retain the pixel intensity and detailed texture information of input images, the reconstruction loss function is given by
\begin{equation}\label{L2}
{L_2} = {\alpha_2}f(I,\hat I) + {\alpha _3}f(V,\hat V) + {\alpha _4}{\left\| {\nabla V - \nabla \hat V} \right\|_1},
\end{equation}
where $I$ and $\hat I$, $V$ and $\hat V$ represent the input and reconstructed images of infrared and visible images, respectively. $\nabla$ denotes the gradient operator, and
\begin{equation}\label{fxxh}
f(X,\hat X) = \left\| {X - \hat X} \right\|_2^2 + \lambda {L_{SSIM}}(X,\hat X),
\end{equation}
where $X$ and $\hat X$ represent the above input image and the reconstructed image, and $\lambda$ is the hyperparameter. SSIM is the structural similarity index\cite{wang2004image}, which is a measure of the similarity between two pictures. Then $L_{SSIM}$ can be described as \[{L_{SSIM}}(X,\hat X) = \frac{{1 - SSIM(X,\hat X)}}{2}.\]

Remark that $L_2$-norm measures the pixel intensity agreement between original and reconstructed images, and that $L_{SSIM}$ computes image dissimilarity in terms of brightness, contrast and structure. Specially, since visible images are with enriched textures, the reconstruction of visible images is regularized by gradient sparsity penalty to guarantee texture agreement.

Combining Eqs.~(\ref{L1}) and (\ref{L2}), the total loss $L_{total}$ can be expressed as
\begin{equation}\label{totalloss}
\begin{split}
  L_{total} =& L_1+L_2 \\
    =& \Phi \left( {\left\| {{B_V} - {B_I}} \right\|_2^2} \right) - {\alpha _1}\Phi \left( {\left\| {{D_V} - {D_I}} \right\|_2^2} \right) \\
    +& {\alpha _2}f(I,\hat I) + {\alpha _3}f(V,\hat V) + {\alpha _4}{\left\| {\nabla V - \nabla \hat V} \right\|_1},
\end{split}
\end{equation}
where ${\alpha _1},{\alpha _2},{\alpha _3},{\alpha _4}$ are the tuning parameters.

\subsection{Fusion Strategy}\label{ADS}
In the above subsections, we have proposed network structure and loss function. After training, we will acquire a decomposer (or say, encoder) and a decoder. In the test phase, we aim to fuse infrared and visible images. The workflow is shown in Figure\ref{DIDF}(b). Different from training, a fusion layer is inserted in the test phase. It fuses background and detail feature maps separately. In formula, there is
\begin{equation}
	B_F = {\rm Fusion}(B_I,B_V), D_F = {\rm Fusion}(D_I,D_V),
\end{equation}
where $B_F$ and $D_F$ denote the fused background and detail feature maps, respectively. In this paper, three fusion strategies are considered as follows:
\begin{itemize}
  \item Summation method:
  $B_F = {B_I} \oplus {B_V}, D_F = {D_I} \oplus {D_V},$
  where the symbol $\oplus$ means element-wise addition.
  \item Weighted average method:
  $B_F = {\gamma _1}{B_I} \oplus {\gamma _2}{B_V}, D_F = {\gamma _3}{D_I} \oplus {\gamma _4}{D_V},$
  where ${\gamma _1}+{\gamma _2={\gamma _3}+{\gamma _4}=1}$ and the default settings for ${\gamma _i}(i=1,\cdot\cdot\cdot,4)$ are all equal to 0.5.
  \item $L_1$-norm method:
  Referring to \cite{li2018densefuse}, we use the $L_1$-norm as a measure of activity, combining with the softmax operator.
  In detail, we can obtain the activity level map of the fused background and detail feature maps by ${\left\| {{B_i}(x,y)} \right\|}_1$ and ${\left\| {{D_i}(x,y)} \right\|}_1 (i = 1,2)$, where $B_1$, $B_2$, $D_1$ and $D_2$ represent $B_I$, $B_V$, $D_I$ and $D_V$, and $(x,y)$ represents the corresponding coordinates of the feature maps and the fused feature map.
Then the adding weights can be calculated by:
\begin{equation*}\label{}
\begin{split}
\eta _i^B(x,y) = &  \frac{{\psi \left( {{{\left\| {{B_i}(x,y)} \right\|}_1}} \right)}}{{\sum\nolimits_{i = 1}^2 {\psi \left( {{{\left\| {{B_i}(x,y)} \right\|}_1}} \right)} }} , \\
\eta _i^D(x,y) = &  \frac{{\psi \left( {{{\left\| {{D_i}(x,y)} \right\|}_1}} \right)}}{{\sum\nolimits_{i = 1}^2 {\psi \left( {{{\left\| {{D_i}(x,y)} \right\|}_1}} \right)} }},
\end{split}
\end{equation*}
where $\psi (\cdot)$ is a $3\times3$ box blur (also known as a mean filter operator). Consequently, we have
\begin{equation*}\label{}
\begin{split}
  B_F & = (\eta _1^B\otimes{B_I}) \oplus (\eta _2^B\otimes{B_V}), \\
    D_F& = (\eta _1^D\otimes{D_I})\oplus (\eta _2^D\otimes{D_V}).
\end{split}
\end{equation*}
where $\otimes$ means element-wise multiplication.
\end{itemize}
\section{Experiment} \label{sec:4}
The aim of this section is to study the performance of our proposed model and compare it with other SOTA models, including FusionGAN~\cite{ma2019fusiongan}, Densefuse~\cite{li2018densefuse}, ImageFuse~\cite{li2018infrared}, DeepFuse~\cite{prabhakar2017deepfuse}, TSIFVS~\cite{bavirisetti2016two}, TVADMM~\cite{guo2017infrared},
CSR~\cite{liu2016image} and ADF~\cite{bavirisetti2015fusion}. All experiments were conducted with Pytorch on a computer with Intel Core i7-9750H CPU@2.60GHz and RTX2070 GPU.

We employ six metrics to evaluate the quality of a fused image, that is,
entropy (EN), mutual information (MI), standard deviation (SD), spatial frequency (SF), visual information fidelity (VIF) and average gradient (AG). More details for these metrics can be seen in \cite{ma2019infrared}.

\paragraph{Datasets and preprocessing.}Our experiments are conducted on three datasets, including TNO \cite{TNO}, NIR \cite{brown2011multi} and FLIR (available at \url{https://github.com/jiayi-ma/RoadScene}).
In our experiment, we divide them into training, validation, and test sets. Table~\ref{dataset} shows the numbers of image pairs, illumination and scene information of the datasets.
We randomly selected 180 pairs of images in the FLIR dataset as training samples. Before training, all images are transformed into grayscale. At the same time, we center-crop them with $128\times128$ pixels.
\begin{table}[tbp]
\centering
\begin{tabular}{cccc}
\toprule
	&	Dataset(pairs)	&	Illumination	\\
\midrule
Training	&	FLIR-Train(180)	&	Daylight\&Nightlight	\\
\midrule
\multirow{2}*{Validation} 	&	NIR-Urban(58)		&	Daylight	\\
	&	NIR-Street(50)	&	Daylight	\\
\midrule
\multirow{3}*{Test} 	&	TNO	(40)	&	Nightlight	\\
	&	FLIR-Test(40)	&	Daylight\&Nightlight	\\
	&	NIR-Country(52)	&	Daylight	\\
\bottomrule
\end{tabular}
\caption{Dataset used in this paper.}
\label{dataset}
\end{table}

\begin{figure}[tb]
	\centering
	\includegraphics[width=1\linewidth]{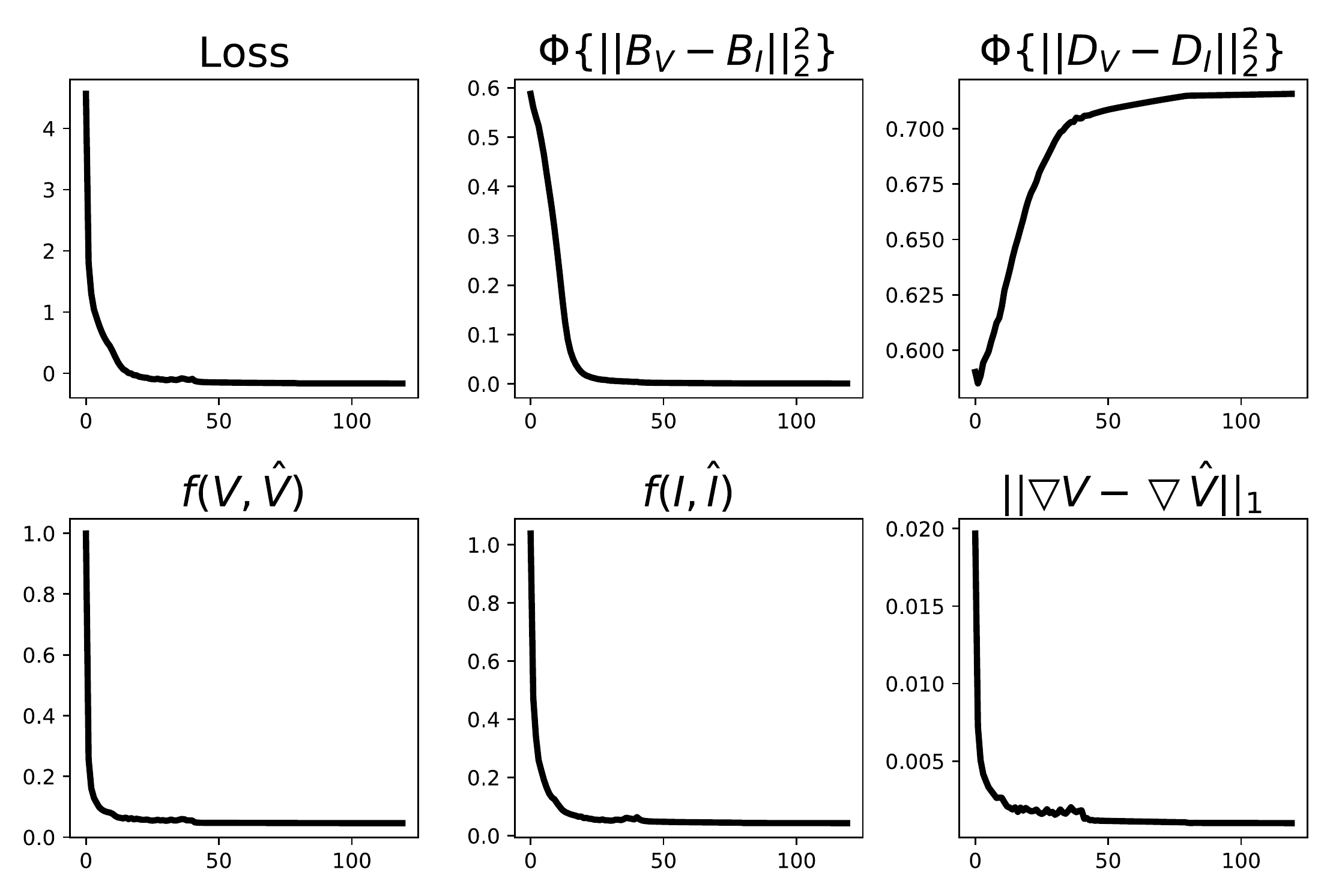}
	\caption{Loss curves over 120 epochs.}
	\label{Loss}
\end{figure}

\paragraph{Hyperparameters setting.}The tuning parameters in loss function are empirically set as follows:  ${\alpha _1}=0.05$, ${\alpha _2}= 2$, ${\alpha _3}=2$, ${\alpha _4}=10$ and $\lambda=5$. In training phase, the network is optimized by Adam over 120 epochs with a batch size of 24.
As for learning rate, we set it to $10^{-3}$ and decrease it by 10 times every 40 epochs.
Figure \ref{Loss} displays loss curves versus epoch index. It is shown that all loss curves are very flat after 120 epochs. In other words, the network is able to converge with this configuration.

\subsection{Experiments on Fusion Strategy}\label{Va}
As described in section \ref{ADS}, fusion strategy plays an important role in our model. We investigate the performance of three strategies on validation set. Table \ref{table_VAL} reports numerical results of six metrics on validation set. Obviously, it is shown that summation strategy achieves higher values, especially in terms of SD, SF, VIF and AG. Hence, the following experiments
adopt summation strategy.

\begin{table}[tbp]
	\centering
	\begin{tabular}{crrr}
		\toprule
		\multicolumn{4}{c}{\textbf{Dataset: NIR Dataset. Scene: Street
		}}\\
		Method&\makebox[1.5cm][c]{Summation}&\makebox[1.5cm][c]{Average}&\makebox[1.5cm][c]{$L_1$-norm}\\
		\midrule
		EN	&	\textbf{7.17 	$\pm$	0.10} 	&	6.85 	$\pm$	0.03 	&	6.87 	$\pm$	0.03 	\\
		MI	&	\textbf{4.69 	$\pm$	0.06} 	&	4.68 	$\pm$	0.04 	&	4.68 	$\pm$	0.03 	\\
		SD	&	\textbf{55.51 	$\pm$	1.74} 	&	36.51 	$\pm$	0.59 	&	36.88 	$\pm$	0.65 	\\
		SF	&	\textbf{24.28 	$\pm$	1.12} 	&	16.85 	$\pm$	0.23 	&	16.80 	$\pm$	0.23 	\\
		VIF	&	\textbf{1.02 	$\pm$	0.04} 	&	0.62 	$\pm$	0.01 	&	0.63 	$\pm$	0.01 	\\
		AG	&	\textbf{7.18 	$\pm$	0.41} 	&	4.89 	$\pm$	0.07 	&	4.88 	$\pm$	0.07 	\\
		\midrule													
		\multicolumn{4}{c}{\textbf{Dataset: NIR Dataset. Scene: Urban}}\\
		Method&\makebox[1.5cm][c]{Summation}&\makebox[1.5cm][c]{Average}&\makebox[1.5cm][c]{$L_1$-norm}\\	
		\midrule	
		EN	&	\textbf{7.18 	$\pm$	0.14} 	&	7.12 	$\pm$	0.03 	&	7.12 	$\pm$	0.03 	\\
		MI	&	6.07 	$\pm$	0.07 	&	\textbf{6.15 	$\pm$	0.04} 	&	6.14 	$\pm$	0.03 	\\
		SD	&	\textbf{61.46 	$\pm$	1.55} 	&	41.64 	$\pm$	0.51 	&	41.73 	$\pm$	0.50 	\\
		SF	&	\textbf{29.22 	$\pm$	1.14} 	&	20.20 	$\pm$	0.22 	&	20.17 	$\pm$	0.22 	\\
		VIF	&	\textbf{1.13 	$\pm$	0.05} 	&	0.77 	$\pm$	0.01 	&	0.77 	$\pm$	0.01 	\\
		AG	&	\textbf{8.13 	$\pm$	0.41} 	&	5.85 	$\pm$	0.06 	&	5.85 	$\pm$	0.06 	\\
		\bottomrule

	\end{tabular}
		  \caption{Results of validation set for choosing the addition strategy.}
\label{table_VAL}
\end{table}

\subsection{Experiments on Image Decomposition}
\begin{figure}[tb]
	\centering
	\includegraphics[width=\linewidth]{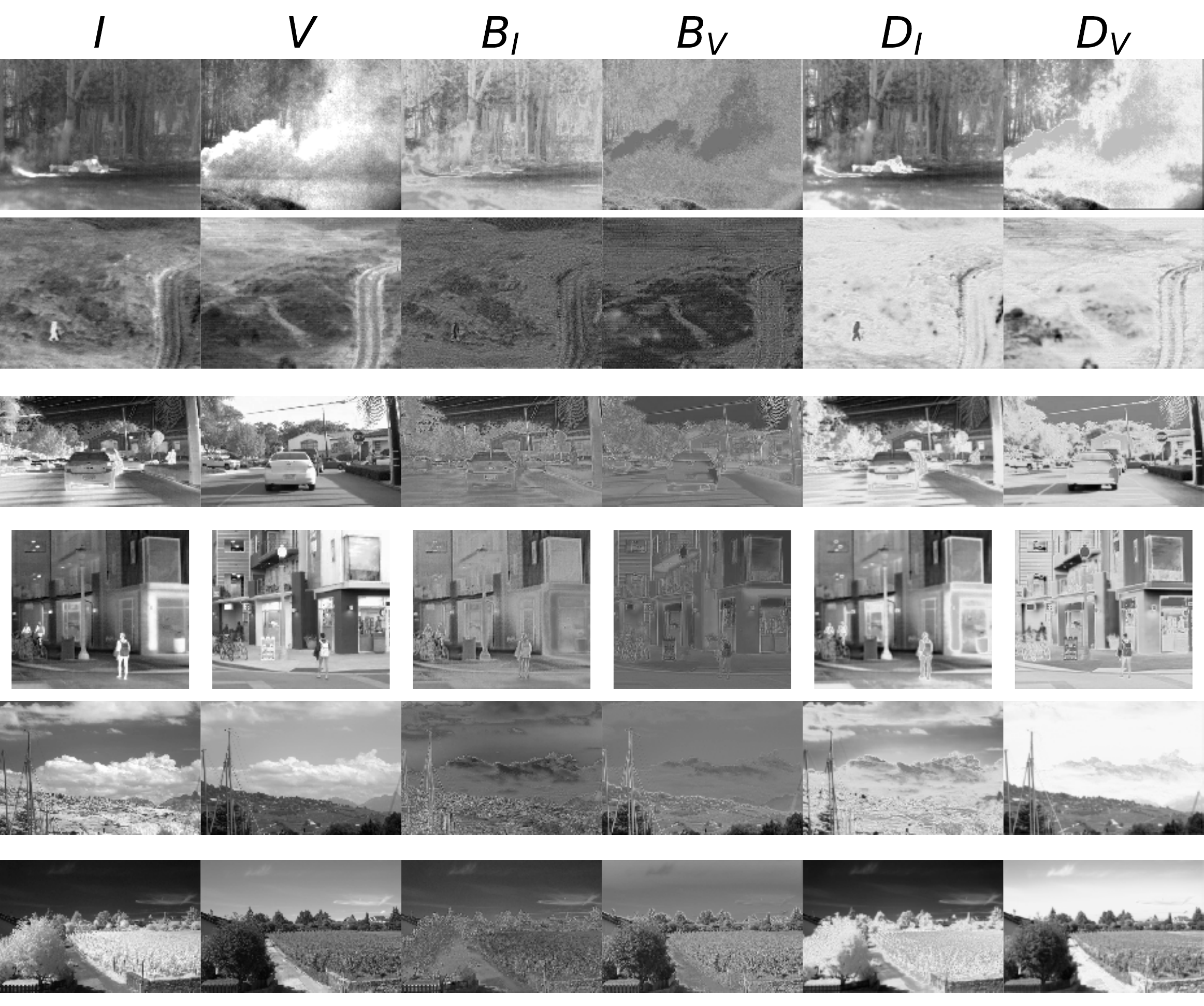}
	\caption{Illustration of deep image decomposition. From left to right: infrared image, visible image, background and detail feature maps of infrared image and visible image. }
	\label{base_detail}
\end{figure}

One of our contributions is the deep image decomposition. It is interesting to study whether decomposed feature maps are able to meet our demands. In Figure \ref{base_detail}, it displays the first channels of feature maps which are generated by \texttt{conv3} and \texttt{conv4}. It is evident that our method can separate the backgrounds and details of infrared and visible images. For background feature maps, it is found that $B_I$ and $B_V$ are visually similar, and they reflect the background and environment of the same scene. Conversely, the gap between $D_I$ and $D_V$ is large, which illustrates the distinct characteristics contained in different source images. That is, the infrared images contain target highlight and thermal radiation information while gradient and texture information of targets are involved in the visible images. In conclusion, it to some degree verifies the rationality of our proposed network structure and image decomposition loss function.

\subsection{Comparison with Other Models}
In this subsection, we will compare our model with the other popular counterparts.

\begin{figure}[!]
	\centering
	\includegraphics[width=\linewidth]{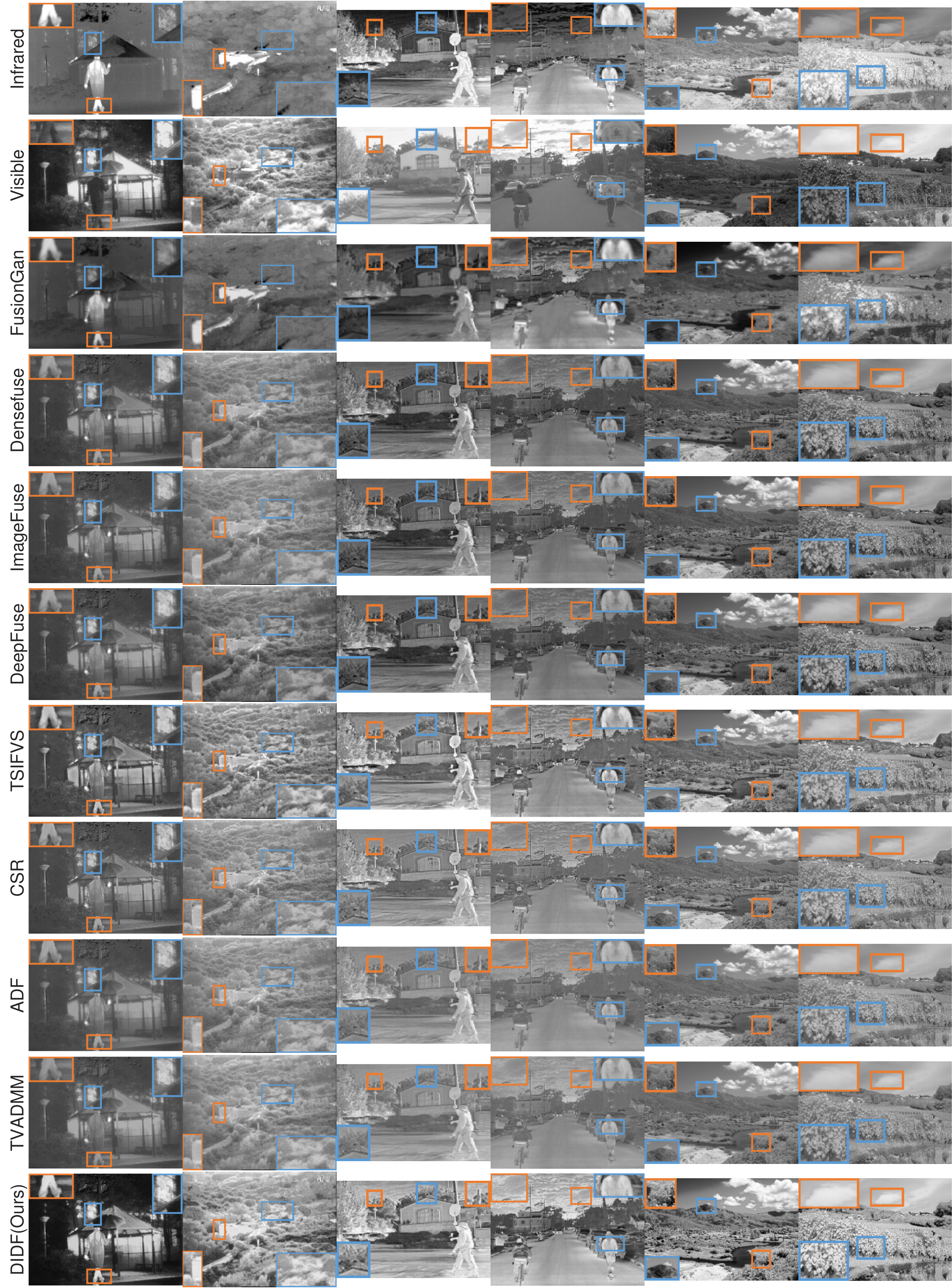}
	\caption{Qualitative results for different methods. Areas marked by orange and blue boxes are amplified for ease of inspection.}
	\label{Figure}
\end{figure}

\paragraph{Qualitative comparison.} Figure~\ref{Figure} exhibits several representative fusion images generated by different models.
Visual inspection shows that, in the images containing people, other methods have problems such as weak high-lighted objects, poor contrast and less prominent contour of targets and backgrounds.
Similarly, if the images are natural landscapes, others have blurred boundaries of mountains and trees, poor color contrast, and insufficient sharpness.
Conversely, our method can obtain fused images with brighter targets, sharper edge contours and retaining richer detailed information.

\paragraph{Quantitative comparison.}\begin{table*}[h]
	\centering
\begin{tabular}{lccccccccc}
\toprule
 \multicolumn{10}{c}{\textbf{Dataset: TNO image fusion dataset}}\\
Metrics	&	FusionGAN	&	DenseFuse	&	ImageFuse	&	DeepFuse	&	TSIFVS	&	TVADMM	&	CSR	&	ADF	&		DIDFuse\\ \midrule
EN	&	6.576 	&	6.842 	&	6.382 	&	\underline{6.860}	&	6.669 	&	6.402 	&	6.428 	&	6.399 	&	\textbf{7.006} 	\\
MI	&	\underline{2.341} 	&	2.302 	&	2.155 	&	2.298 	&	1.717 	&	2.041 	&	1.990 	&	2.007 	&	\textbf{2.347} 	\\
SD	&	29.035 	&	31.817 	&	22.938 	&	\underline{32.249} 	&	28.036 	&	23.007 	&	23.603 	&	22.963 	&	\textbf{42.657} 	\\
SF	&	8.762 	&	11.095 	&	9.800 	&	11.125 	&	12.598 	&	9.034 	&	\underline{11.445} 	&	10.782 	&	\textbf{13.126} 	\\
VIF	&	0.258 	&	0.572 	&	0.306 	&	\underline{0.581} 	&	0.456 	&	0.284 	&	0.312 	&	0.286 	&	\textbf{0.623} 	\\
AG	&	2.417 	&	3.597 	&	2.719 	&	3.599 	&	\underline{3.980} 	&	2.518 	&	3.367 	&	2.988 	&	\textbf{4.294} 	\\
 \midrule
 \multicolumn{10}{c}{\textbf{Dataset: FLIR image fusion dataset}}\\
Metrics	&	FusionGAN	&	DenseFuse	&	ImageFuse	&	DeepFuse	&	TSIFVS	&	TVADMM	&	CSR	&	ADF	&		DIDFuse	\\\midrule
EN	&	7.017 	&	7.213 	&	6.992 	&	\underline{7.213 }	&	7.152 	&	6.797 	&	6.909 	&	6.798 	&	\textbf{7.344} 	\\
MI	&	2.684 	&	2.727 	&	\underline{2.783} 	&	2.725 	&	2.312 	&	2.473 	&	2.569 	&	2.721 	&	\textbf{2.882} 	\\
SD	&	34.383 	&	37.315 	&	32.579 	&	\underline{37.351} 	&	35.889 	&	28.071 	&	30.529 	&	28.371 	&	\textbf{46.888} 	\\
SF	&	11.507 	&	15.496 	&	14.519 	&	15.471 	&	\underline{18.794} 	&	14.044 	&	17.128 	&	14.480 	&	\textbf{18.835} 	\\
VIF	&	0.289 	&	0.498 	&	0.419 	&	0.498 	&	\underline{0.503} 	&	0.325 	&	0.373 	&	0.337 	&	\textbf{0.538} 	\\
AG	&	3.205 	&	4.822 	&	4.150 	&	4.802 	&	\underline{5.568} 	&	3.524 	&	4.799 	&	3.564 	&	\textbf{5.574} 	\\
 \midrule
 \multicolumn{10}{c}{\textbf{Dataset: RGB-NIR Scene Dataset}}\\
Metrics	&	FusionGAN	&	DenseFuse	&	ImageFuse	&	DeepFuse	&	TSIFVS	&	TVADMM	&	CSR	&	ADF	&	 DIDFuse 	\\\midrule
EN	&	7.055 	&	\underline{7.304} 	&	7.217 	&	7.303 	&	7.300 	&	7.129 	&	7.170 	&	7.105 	&	\textbf{7.357} 	\\
MI	&	3.003 	&	\textbf{4.045} 	&	3.967 	&	\underline{4.040}	&	3.285 	&	3.673 	&	3.699 	&	3.944 	&	3.795 	\\
SD	&	34.912 	&	\underline{45.850} 	&	42.307 	&	45.815 	&	43.743 	&	40.469 	&	40.383 	&	38.978 	&	\textbf{57.557} 	\\
SF	&	14.309 	&	18.718 	&	18.360 	&	18.627 	&	\underline{20.646} 	&	16.685 	&	20.370 	&	17.313 	&	\textbf{26.529} 	\\
VIF	&	0.424 	&	0.677 	&	0.613 	&	0.676 	&	\underline{0.688} 	&	0.530 	&	0.583 	&	0.538 	&	\textbf{0.937} 	\\
AG	&	4.564 	&	6.228 	&	5.920 	&	6.178 	&	\underline{6.823} 	&	5.319 	&	6.488 	&	5.381 	&	\textbf{8.741} 	\\
\bottomrule
\end{tabular}
    \caption{Quantitative results of different methods. The largest value is shown in bold, and the second largest value is underlined.}
    \label{table}
\end{table*}
Subsequently, quantitative comparison results on test set are listed in Table \ref{table}. It is found that our model is the best performer on all datasets in terms of all metrics. As for competitors, they may perform well on a dataset in terms of part of metrics. This result demonstrates that images fused by our model are with enriched textures and satisfy human visual system.
\subsection{Experiments on Reproducibility}
\begin{figure}[h]
	\centering	\includegraphics[width=\linewidth]{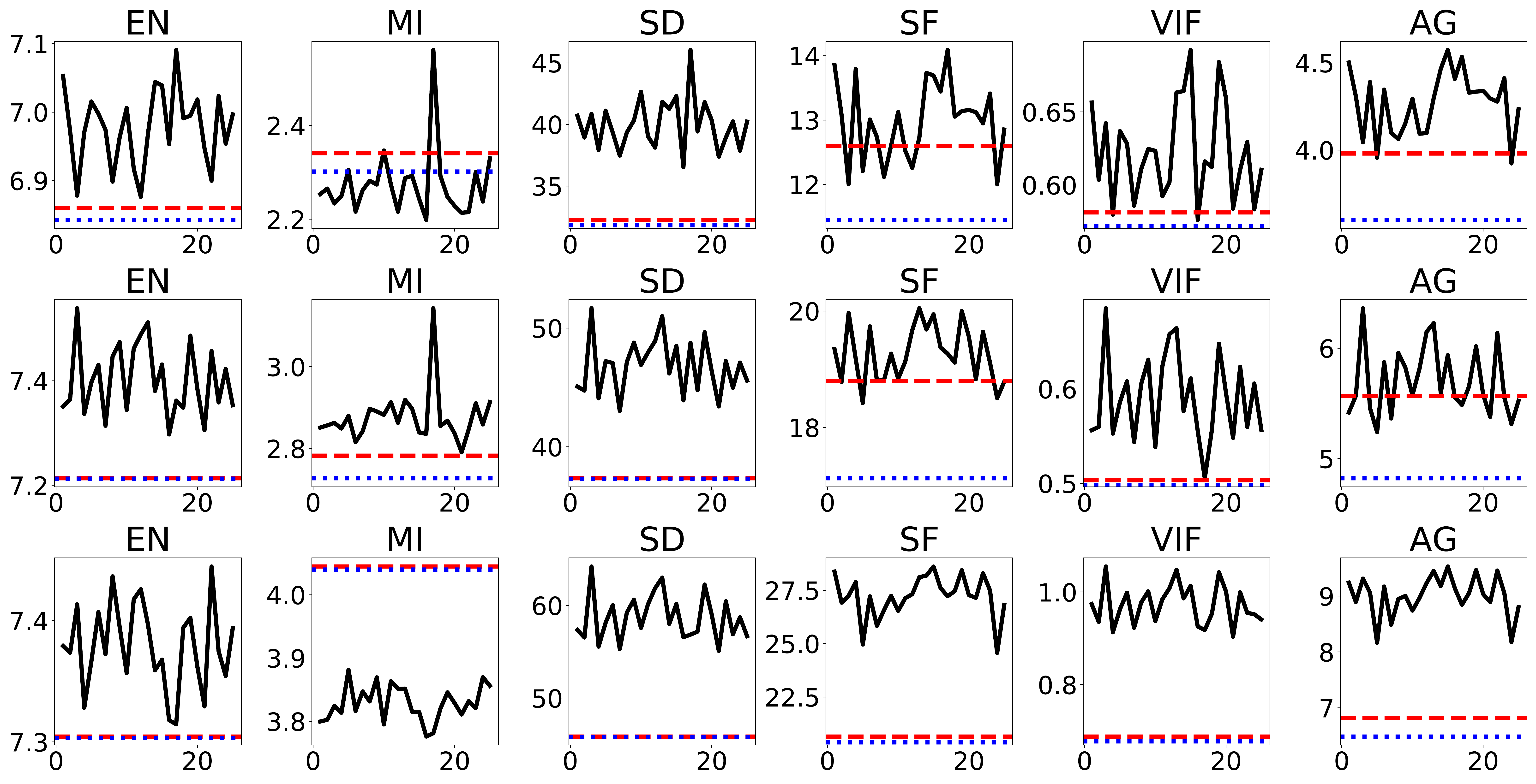}
	\caption{Test results of model reproducibility. From top to bottom: image fusion dataset TNO, FLIR, and NIR. From left to right: the values of EN, MI, SD, SF, VIF and AG.}
	\label{Robustness}
\end{figure}

As is known, deep learning methods are often criticized for instability. Therefore, we test the reproducibility of DIDFuse in the last experiment. We repeatedly train the network 25 times and quantitatively compare the 25 parallel results. As shown in Figure \ref{Robustness}, the black solid curves report six metrics over 25 experiments. The red dashed line and blue dotted line represent the greatest and the second greatest values in the comparison methods, respectively.
Similar to the above results, our method can basically keep the first place all the time, indicating that DIDFuse can generate high-quality fused images steadily.

\section{Conclusion}\label{sec:5}
To solve the IVIF issue, we construct a new AE network in which the encoder is used for two-scale image decomposition and the decoder is responsible for image reconstruction.
In the training phase, the encoder is trained to output background and feature maps, then the decoder reconstructs original images.
In the test phase, we set a fusion layer between the encoder and decoder to fuse background and detail feature maps through a specific fusion strategy. Finally, the fused image can be acquired through the decoder.
We test our model on TNO, FLIR, and NIR datasets. Qualitative and quantitative results show that our model outperforms other SOTA methods, since our model can steadily obtain a fusion image of highlighted targets and rich details.

\section*{Acknowledgments}
The research is supported by the National Key Research and Development Program of China under grant 2018AAA0102201 and 2018YFC0809001,
the National Natural Science Foundation of China under grant 61976174, 11671317 and 61877049, the Fundamental Research Funds for the Central Universities under grant number xzy022019059.

\newpage
\bibliographystyle{named}
\bibliography{xbib}
\end{document}